\documentclass[apj]{emulateapj}			%

\usepackage{epsfig}
\usepackage{natbib}                
\usepackage{verbatim}              
\usepackage{graphicx}              
\usepackage{amsmath,amsthm,amsfonts,amsopn,amssymb} 
\usepackage{afterpage}
\usepackage{rotating}					

\shorttitle{55~Cnc: Stellar Parameters, Transiting Super-Earth, and Habitable Zone Planet}

\shortauthors{von Braun \& Boyajian et al.}
\begin{document}


\title{55 Cancri: Stellar Astrophysical Parameters, a Planet in the Habitable Zone, and Implications for the Radius of a Transiting Super-Earth} 

\author{Kaspar von Braun\altaffilmark{1,11},
Tabetha S. Boyajian\altaffilmark{2,3}, 
Theo A. ten Brummelaar\altaffilmark{6}, 
Stephen R. Kane\altaffilmark{1},
Gerard T. van Belle\altaffilmark{4},  
David R. Ciardi\altaffilmark{1},  
Sean N. Raymond\altaffilmark{9,10},
Mercedes L\'{o}pez-Morales\altaffilmark{5,8},
Harold A. McAlister\altaffilmark{2}, 
Gail Schaefer\altaffilmark{6},
Stephen T. Ridgway\altaffilmark{7}, 
Laszlo Sturmann\altaffilmark{6}, 
Judit Sturmann\altaffilmark{6}, 
Russel White\altaffilmark{6},
Nils H. Turner\altaffilmark{6},
Chris Farrington\altaffilmark{6}, 
and P. J. Goldfinger\altaffilmark{6} 
}

\altaffiltext{1}{NASA Exoplanet Science Institute, California Institute of Technology, MC 100-22, Pasadena, CA 91125}
\altaffiltext{2}{Center for High Angular Resolution Astronomy and Department of Physics and Astronomy, Georgia State University, P. O. Box 4106, Atlanta, GA 30302-4106} 
\altaffiltext{3}{Hubble Fellow} 
\altaffiltext{4}{European Southern Observatory, Karl-Schwarzschild-Str. 2, 85748 Garching, Germany}
\altaffiltext{5}{Institut de Ci\`{e}ncies de L'Espai (CSIC-IEEC), Campus UAB, Facultat Ci\`{e}ncies, Torre C5 parell 2, 08193 Bellaterra, Barcelona, Spain}
\altaffiltext{6}{The CHARA Array, Mount Wilson Observatory, Mount Wilson, CA 91023}
\altaffiltext{7}{National Optical Astronomy Observatory, P.O. Box 26732, Tucson, AZ 85726-6732}
\altaffiltext{8}{Department of Terrestrial Magnetism, Carnegie Institution of Washington, 5241 Broad Branch Road, NW, Washington, DC 20015}
\altaffiltext{9}{Universit{\'e} de Bordeaux, Observatoire Aquitain des Sciences de l'Univers, 2 rue de l'Observatoire, BP 89, F-33271 Floirac Cedex, France}
\altaffiltext{10}{CNRS, UMR 5804, Laboratoire d'Astrophysique de Bordeaux, 2 rue de l'Observatoire, BP 89, F-33271 Floirac Cedex, France}
\altaffiltext{11}{kaspar@caltech.edu}



\begin{abstract}

The bright star 55~Cancri is known to host five planets, including a transiting super-Earth. The  study presented here yields directly determined values for 55~Cnc's stellar astrophysical parameters based on improved interferometry: $R=0.943 \pm 0.010 R_{\odot}$, $T_{\rm EFF} = 5196 \pm 24$ K. We use isochrone fitting to determine 55~Cnc's age to be 10.2 $\pm$ 2.5 Gyr, implying a stellar mass of $0.905 \pm 0.015 M_{\odot}$. Our analysis of the location and extent of the system's habitable zone (0.67--1.32 AU) shows that planet f, with period $\sim$ 260 days and $M \sin i = 0.155 M_{Jupiter}$, spends the majority of the duration of its elliptical orbit in the circumstellar habitable zone. Though planet f is too massive to harbor liquid water on any planetary surface, we elaborate on the potential of alternative low-mass objects in planet f's vicinity: a large moon, and a low-mass planet on a dynamically stable orbit within the habitable zone. Finally, our direct value for 55~Cancri's stellar radius allows for a model-independent calculation of the physical diameter of the transiting super-Earth 55~Cnc e ($\sim 2.05 \pm 0.15 R_{\earth}$), which, depending on the planetary mass assumed, implies a bulk density of 0.76 $\rho_{\earth}$ or 1.07 $\rho_{\earth}$. 

\end{abstract}



\section{Introduction}\label{sec:introduction}


55 Cancri (= HD~75732 = $\rho$ Cancri; 55~Cnc hereafter) is a late G / early K dwarf / subgiant \citep{gra03} currently known to host five extrasolar planets with periods between around 0.7 days and 14 years and minimum masses between 0.026 and 3.84 $M_{Jupiter}$ \citep{daw10}. These planets were all discovered via the radial velocity method and successively announced in \citet{but97}, \citet{mar02}, \citet{mca04}, and \citet{fis08}. 

Astrophysical insights on two of the currently known planets in the 55~Cnc system are direct consequences of the determination of the stellar radius and surface temperature: 

\begin{itemize}

\item Elimination of period aliasing and the consequently updated orbital scenario presented in \citet{daw10} motivated the recent, successful photometric transit detections of the super-Earth 55~Cnc e with a period of 0.7 days by \citet{win11} using $MOST$ and, independently, \citet{dem11} using Warm $Spitzer$. The calculation of planetary radii based on transit photometry relies, of course, on a measured or assumed stellar radius. 



\item Based on the equations relating stellar luminosity to the location and extent of a stellar system's  habitable zone (HZ) \citep{jon10}, planet 55~Cnc f falls within 55~Cnc's traditional circumstellar HZ.

\end{itemize}

Apart from values derived from stellar modeling \citep{fis08}, there are two direct (interferometric) stellar diameter determinations of 55~Cnc: $R = 1.15 \pm 0.035 R_{\odot}$ in \citet{bai08} and $R = 1.1 \pm 0.096 R_{\odot}$ in \citet{van09}. Note, however, that \citet{van09} report, in their \S 5.1 and \S 5.4.1, the fact that $R \sim 1.1 R_{\odot}$ makes 55~Cnc a statistical outlier in their fitted $T_{\rm EFF} = f((V-K)_0)$ relation (see their \S 5.1). In order not to be an outlier, 55~Cnc's angular diameter would have to be 0.7 milliarcseconds (mas), corresponding to 0.94$R_{\odot}$.  


In this paper, we present new, high-precision interferometric observations of 55~Cnc with the aim of providing a timely, directly determined value of the stellar diameter, which, when combined with the flux decrement measured during planetary transit, yields a direct value for the exoplanetary diameter. Furthermore, the combination of angular stellar diameter and bolometric stellar flux provides directly determined stellar surface temperature. The resultant stellar luminosity determines the location and extent of the circumstellar HZ, and we can ascertain which, if any, of the planets orbiting 55~Cnc spend any, all, or part of their orbits inside the HZ. 


We describe our observations in \S \ref{sec:observations}. The determination of stellar astrophysical parameters is shown in \S \ref{sec:properties}. We discuss 55~Cnc's circumstellar HZ and the locations of the orbiting planets with respect to it in \S \ref{sec:hz}. Section \ref{sec:planet} contains the  calculation of the radius of the transiting super-Earth 55~Cnc e, and we conclude in \S \ref{sec:conclusion}.

\section{Interferometric Observations}\label{sec:observations}



Our observational strategy is described in detail in \citet{von11}. We briefly repeat the general approach here.

55~Cnc was observed on the nights of 11 and 12 May, 2011, using the Georgia State University Center for High Angular Resolution Astronomy (CHARA) Array \citep{ten05}, a long baseline interferometer located at Mount Wilson Observatory in Southern California. We used the CHARA Classic beam combiner with CHARA's longest baseline, S1E1 ($\sim$ 330~m) to collect the observations in $H$-band ($\lambda_{central} = 1.67$~$\mu$m).

The interferometric observations included the common technique of taking bracketed sequences of the object with calibrator stars, designed to characterize and subsequently eliminate the temporally variable effects of the atmosphere and telescope/instrument upon our calculation of interferometric visibilities\footnote{Visibility is the normalized amplitude of the correlation of the light from two telescopes. It is a unitless number ranging from 0 to 1, where 0 implies no correlation, and 1 implies perfect correlation. An unresolved source would have perfect correlation of 1.0 independent of the distance between the telescopes (baseline). A resolved object will show a decrease in visibility with increasing baseline length. The shape of the visibility versus baseline is a function of the topology of the observed object (the Fourier Transform of the object's shape). For a uniform disk this function is a Bessel function, and for this paper, we use a simple model of a limb darkened variation of a uniform disk.}. During the observing period, we alternated between two point-source like calibrators, both of which lie within 5 degrees on the sky from the target, to minimize the systematic effects. These calibrator stars were: HD~74811 (G2~IV; $\theta_{EST} = 0.407\pm0.015$~mas) and HD~75332 (F8~V; $\theta_{EST} = 0.401\pm0.014$~mas). $\theta_{EST}$ corresponds the estimated angular diameter of the calibrator stars based on spectral energy distribution fitting. 

\begin{figure*}										
  \begin{center}
    \begin{tabular}{cc}
      \epsfig{file=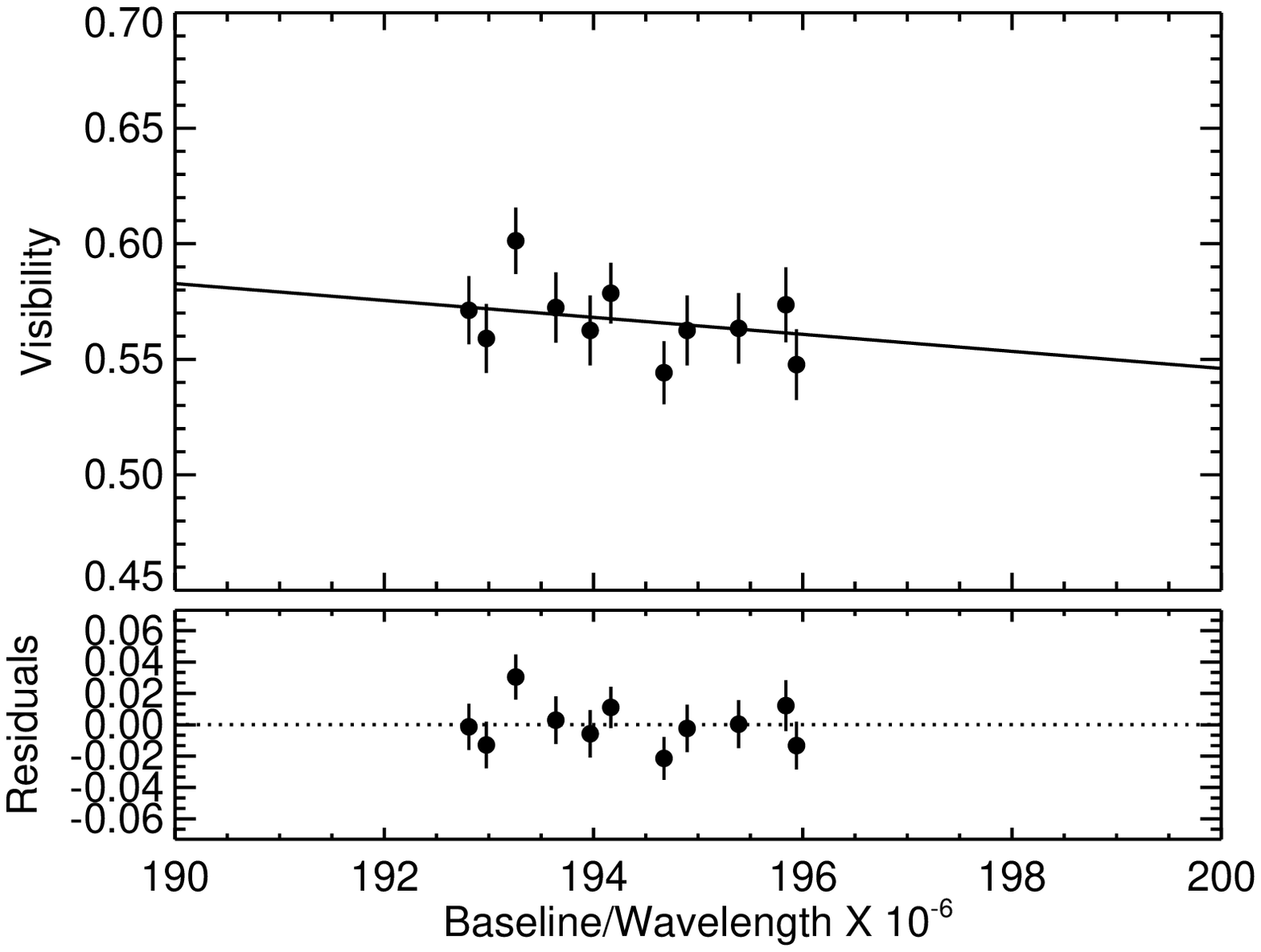,width=0.5\linewidth,clip=} &       
      \epsfig{file=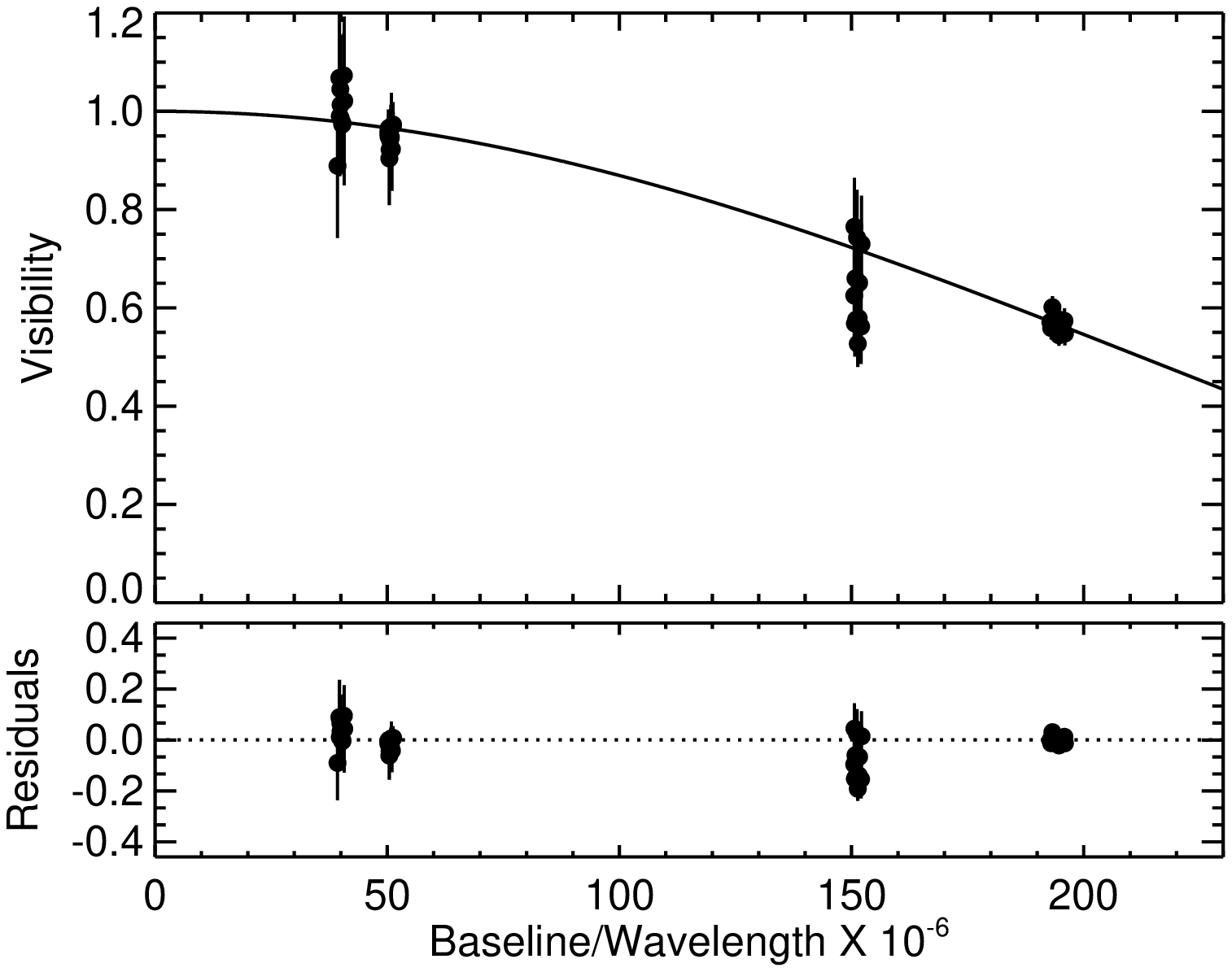,width=0.5\linewidth,clip=} \\
    \end{tabular}
  \end{center}
  \caption{Calibrated visibility observations along with the limb-darkened angular diameter fit for 55~Cnc. The left panel shows the fit only based on the CHARA data presented in this work. For comparison with literature datasets, the right panel includes PTI data (shortest baseline) from \citet{van09}, CHARA data from \citet{bai08} (longer baseline), and data presented in this work (longest baseline), along with the same fit as in the left planel.  The bottom panels show the residuals around the respective fit. Note the different scales for both x and y axes between the panels. For details, see \S \ref{sec:observations} and \S \ref{sec:radius}. The fit results are given in Table \ref{tab:properties}.}
  \label{fig:visibility}
\end{figure*}

The uniform disk and limb-darkened angular diameters $\theta_{\rm UD}$ and $\theta_{\rm LD}$\footnote{The limb-darkening corrected $\theta_{\rm LD}$ corresponds to the angular diameter of the Rosseland, or mean, radiating surface of the star.}, respectively, are found by fitting our calibrated visibility measurements 
to the respective functions for each relation \citep{han74}. Limb darkening coefficients were taken from \citet{cla00}. The data and fit for $\theta_{\rm LD}$ are shown in the left panel of Fig. \ref{fig:visibility}. 


\section{Fundamental Astrophysical Parameters of the Star 55~Cancri}\label{sec:properties}

In this Section, we discuss 55~Cnc's stellar astrophysical properties. The results are summarized in Table \ref{tab:properties}.


\subsection{Stellar Diameter} \label{sec:radius}


We examined the following two sets of literature interferometric data in order to decide whether to include them into our analysis: CHARA data presented in \citet{bai08}, and data published in \citet{van09} taken with the Palomar Testbed Interferometer (PTI), which features a 110~m baseline compared to CHARA's 330~m\footnote{Canonically, CHARA's spatial resolution is therefore superior with respect to PTI data by a factor of three, but PTI data pipeline and its products are well characterized \citep{bod98}.}. In the left panel of Figure \ref{fig:visibility}, we show our data and corresponding fit for $\theta_{\rm LD}$. The right panel of Figure \ref{fig:visibility} contains all three datasets along with the fit based on our new data. The superiority of our new CHARA data due to the longer baselines is readily apparent. We therefore chose to assign zero weight to the two literature datasets in our analysis, particularly due to the fact that fits inclusive of all data weighted equally do not influence the fit and corresponding results.

Our interferometric measurements (Figure \ref{fig:visibility}) yield a limb-darkening corrected angular diameter $\theta_{\rm LD} = 0.711 \pm 0.004$ mas. Combined with 55~Cnc's trigonometric parallax value from \citet{van07}, we calculate its linear radius to be $R = 0.943 \pm 0.010 R_{\rm \odot}$ (Table \ref{tab:properties}). 
 
Our result of $\theta_{\rm LD} \simeq 0.7$ mas is consistent with the PTI-derived value ($\sim 1.1$ mas) published in \citet{van09} at the 1.5$\sigma$ level. Furthermore, $\theta_{\rm LD} \simeq 0.7$ mas exactly corresponds to the angular diameter required for 55~Cnc to fall onto the $T_{\rm EFF}$ versus $(V-K)_0$ relation in \citet{van09}; see their equation 2 and section 5.4.1.
Finally, our directly determined value for 55~Cnc's stellar radius ($R = 0.943 R_{\rm \odot}$) is consistent with the calculated value in \citet{fis08} based on stellar parameters cataloged in \citet{val05}.


\subsection{Stellar Effective Temperature} \label{sec:temp}

Following the procedure outlined in \S3.1 of \citet{vcb07}, we produce a fit of the stellar spectral energy distribution (SED) based on the spectral templates of \citet{pic98} to literature photometry published in \citet{nic57, arg63, mar64, arg66, cow72, ruf76, per77, egg78, ols83, mer86, arr89, gon92, ols93, hau98, cut03, kaz05}; see also the catalog of \citet{gez99}. Typical uncertainties per datum for these photometry data are in the range of 5--8\%.

\begin{figure}										
\centering
\epsfig{file=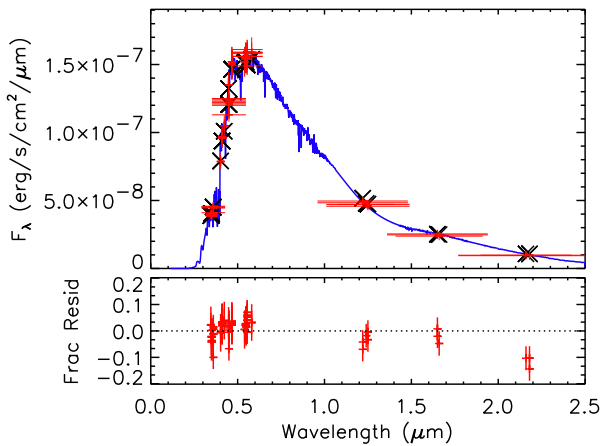,width=1.0\linewidth,clip=} \\
\caption{SED fit for 55~Cnc. The (blue) spectrum is a G9IV spectral template \citep{pic98}. The (red) crosses indicate photometry values from the literature. ``Error bars'' in x-direction represent bandwidths of the filters used. The (black) X-shaped symbols show the flux value of the spectral template integrated over the filter transmission. The lower panel shows the residuals around the fit in fractional flux units of photometric uncertainty. For details, see \S \ref{sec:properties}.}
\label{fig:diameters}
\end{figure}

We obtain fits with reduced $\chi^2 \sim 3$ when using K0~IV and G8~IV spectral templates. The creation of a G9~IV template by linearly interpolating the specific flux values of the K0~IV and G8~IV spectral templates for each value of $\lambda$, however, improves the quality of this fit to a reduced $\chi^2 = 0.72$. 
Interstellar extinction is a free parameter in the fitting process and produces a value of $A_V = 0.000 \pm 0.014$ mag, consistent with expectations for this nearby star. The value for the distance to 55~Cnc is adopted from \citet{van07}. The SED fit for 55~Cnc, along with its residuals, is shown in Fig. \ref{fig:diameters}.

The principal result from the SED fit is the value of 55~Cnc's stellar bolometric flux of $F_{\rm BOL} = (1.227 \pm 0.0177)\times10^{-7}$~erg cm$^{-2}$ s$^{-1}$, and consequently, its luminosity of $L = 0.582 \pm 0.014 L_{\odot}$. Using the rewritten version of the Stefan-Boltzmann Law


\begin{equation} \label{eq:temperature}
T_{\rm EFF} ({\rm K}) = 2341 (F_{\rm BOL}/\theta_{\rm LD}^2)^{\frac{1}{4}},
\end{equation}

\noindent
where $F_{\rm BOL}$ is in units of $10^{-8}$~erg cm$^{-2}$ s$^{-1}$ and  $\theta_{\rm LD}$ is in units of mas, we calculate 55~Cnc's effective temperature to be $T_{\rm EFF} = 5196 \pm 24$ K. 


\subsection{Stellar Mass and Age} \label{sec:age}

\begin{figure*}	
  \begin{center}
      \begin{tabular}{cc}
\epsfig{file=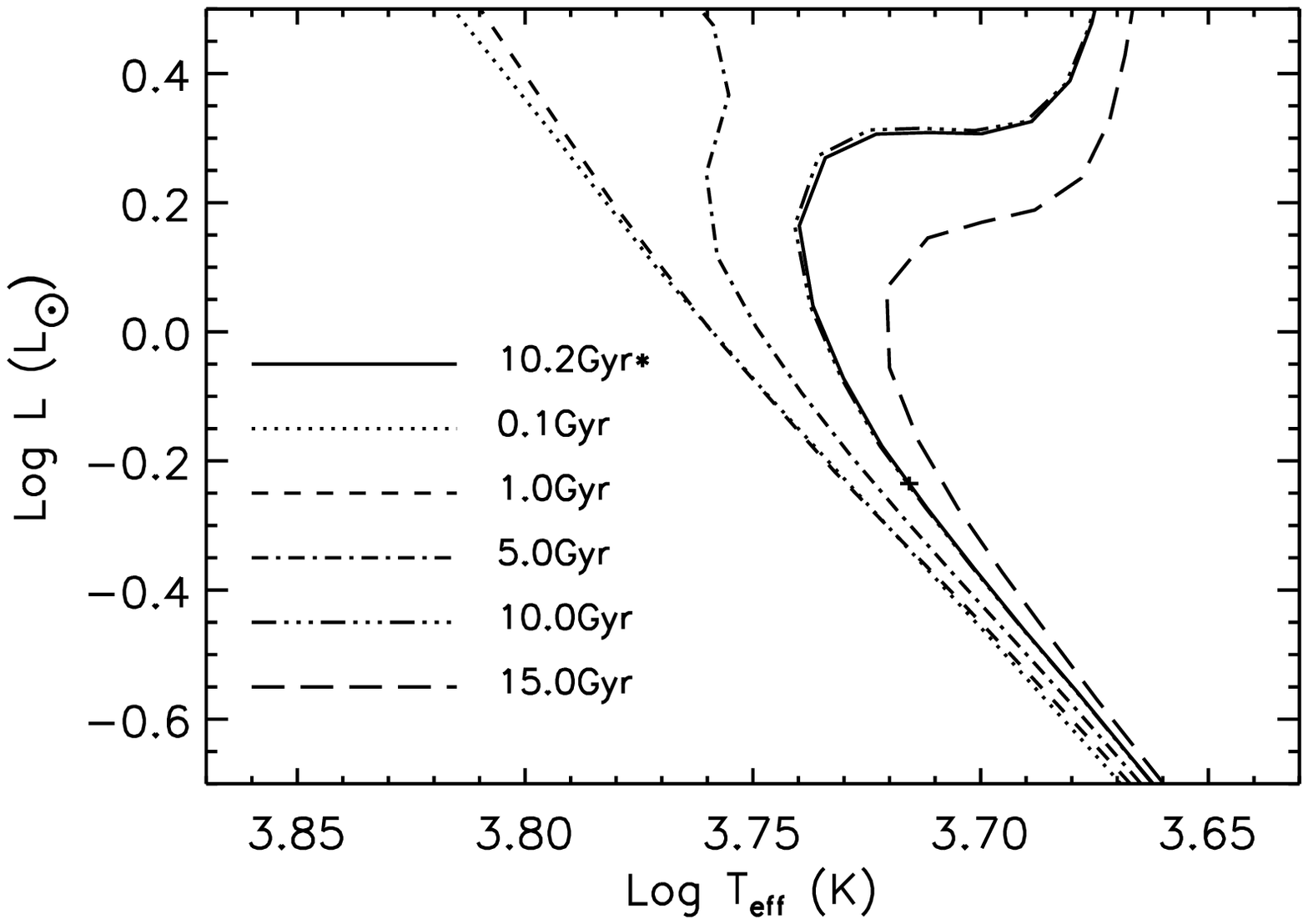,width=0.5\linewidth,clip=} &
\epsfig{file=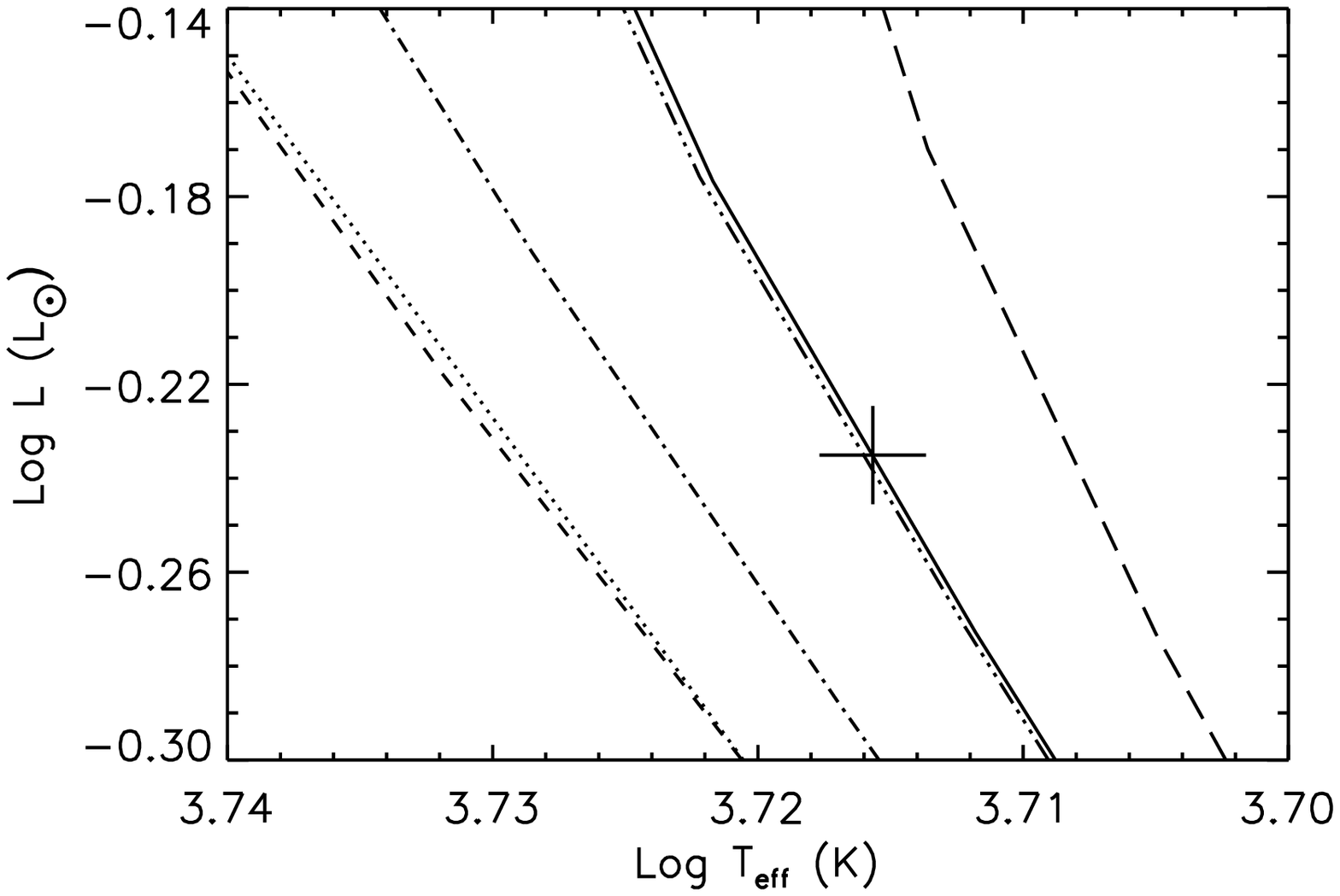,width=0.5\linewidth,clip=} \\
      \end{tabular}
    \end{center}									
\caption{Comparison of our values of 55~Cnc's $L$ and $T_{\rm EFF}$ to Yonsei-Yale isochrones. The right panel is a zoom of the left panel, centered on the data point. Different age isochrones are indicated by the different line styles, where the best fit isochrone (asterisk; solid line) yields a stellar age of 10.2 $\pm$ 2.5 Gyr. For details, see \S \ref{sec:age} and Table~\ref{tab:properties}.}
\label{fig:isochrone}
\end{figure*}

Our values for 55 Cnc's stellar luminosity and effective temperature are compared to  Yonsei-Yale stellar isochrones \citep{dem04,kim02, yi2001} with [Fe/H]~$= 0.31$ \citep{val05,fis08} to estimate its mass and age. As illustrated in Figure~\ref{fig:isochrone}, interpolating between isochrones and mass tracks yields a 
best-fit age of 55~Cnc of 10.2 $\pm$ 2.5~Gyr and stellar mass of 0.905 $\pm$ 0.015$M_{\odot}$.  
Note that the above uncertainties are based on only the 1-$\sigma$ measurement errors in our calculations of $L$ ($\sim$ 2.4\%) and $T_{\rm EFF}$ ($\sim$ 0.5\%), shown as error bars in Figure \ref{fig:isochrone}, and do not take into account systematic offsets due to, e.g., metallicity. 

Our values for 55 Cnc's stellar mass and age are consistent with their respective counterparts obtained via spectroscopic analysis combined with photometric bolometric corrections \citep{val05}, as well as with age estimates using the \ion{Ca}{2} chromospheric activity indicators and gyrochronology relations \citep{wri04, mam08}. In addition, they are within the error bars of the equivalent values derived in \citet{fis08} and the ones used in \citet{win11} and \citet{dem11}.
Finally, the stellar surface gravity is computed from our radius measurement and mass estimate, plus associated uncertainties, to be $\log g = 4.45 \pm 0.01$.  

A summary of all directly determined and calculated stellar astrophysical parameters in this Section is reported in Table~\ref{tab:properties}.


\begin{deluxetable}{rcc}

\tablecaption{Stellar Properties of 55~Cnc \label{tab:properties}} 
\tablewidth{0pc}
\tablehead{
\colhead{Parameter} &
\colhead{Value} &
\colhead{Reference}	
}
\startdata
Spectral Type \dotfill	&	K0 IV-V	&	\citet{gra03} 	\\ 
Parallax (mas) \dotfill				&	$81.03 \pm 0.75$\phn	&	\citet{van07}	\\
$[$Fe/H$]$ \dotfill		&	$0.31 \pm 0.04$	&	\citet{val05}	\\
$\theta_{\rm UD}$ (mas) \dotfill		&	$0.685 \pm 0.004$	&	this work (\S \ref{sec:observations})	\\
$\theta_{\rm LD}$ (mas)	\dotfill	&	$0.711 \pm 0.004$	&	this work	(\S \ref{sec:observations})\\
Radius ($R_{\rm \odot}$) \dotfill	&	$0.943 \pm 0.010$	&	this work	(\S \ref{sec:radius})\\
Luminosity ($L_{\rm \odot}$) \dotfill	& $0.582 \pm 0.014$	&	this work	(\S \ref{sec:temp})\\
$T_{\rm EFF}$ (K)	\dotfill			&	$5196 \pm 24$\phn\phn		&	this work	(\S \ref{sec:temp})\\
Mass ($M_{\rm \odot}$) \dotfill	& 	$0.905 \pm 0.015$	&	this work	(\S \ref{sec:age})\\	
Age (Gyr)	\dotfill	&	$10.2 \pm 2.5$\phn	&	this work	(\S \ref{sec:age})\\	
$\log g$	\dotfill	&	$4.45 \pm 0.01$	&	this work	(\S \ref{sec:age})\\	
HZ boundaries (AU) \dotfill & $0.67 - 1.32$ & this work (\S \ref{sec:hz})\\
\enddata
\tablecomments{Directly determined and derived stellar parameters for the 55~Cnc system.}
\end{deluxetable}


\section{55~Cancri's Habitable Zone}\label{sec:hz}

In this Section, we calculate the location and extent of the HZ in the 55~Cnc system, and we examine which of the orbiting planets spend all or part of their respective orbits in the HZ. We further comment on the potential existence of habitable objects around 55~Cnc.

A circumstellar traditional HZ is defined as the range of distances from a star at which a planet with a moderately dense atmosphere could harbor liquid water on its surface. More details about the definition of habitable zones, typically used for Earth-like planets with clearly defined surfaces, can be found in \citet{kas93} and \citet{und03}. 

\begin{figure*}										
\centering
\epsfig{file=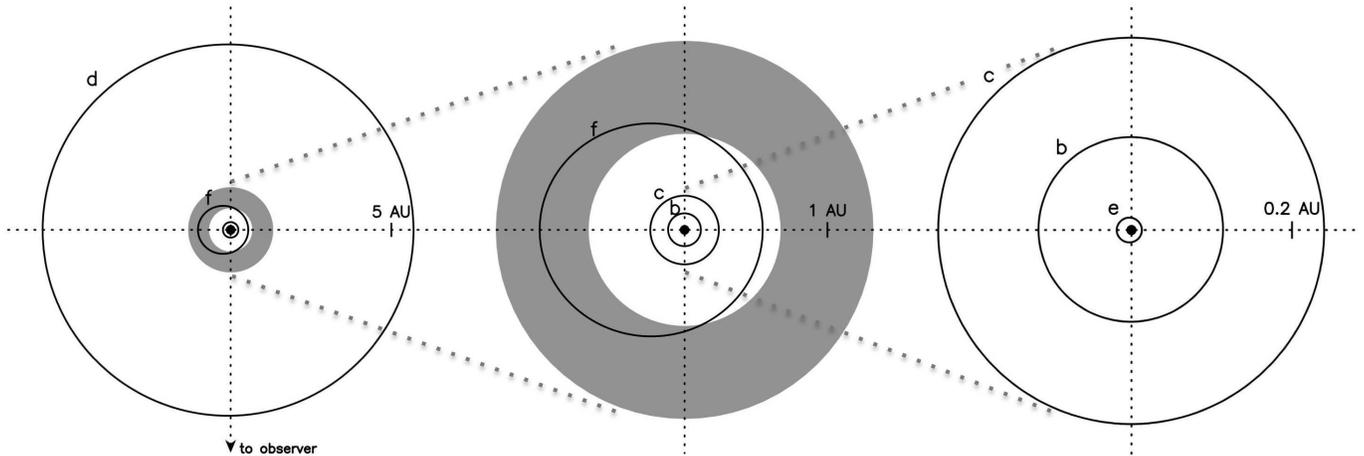,width=1.0\linewidth,clip=} \\
  \caption{A top-down view of the 55~Cnc system, showing the full orbital architecture centered on the star at increasing zoom levels from left to right (note different scales on ordinate). The habitable zone is indicated by the gray shaded region. Orbital element values are adopted from table 10 in \citet{daw10}. Planet d is well beyond the outer edge of the HZ (left panel). Planet f periodically dips into and out of the HZ during its elliptical ($e \simeq 0.3$) orbit (left and middle panels). See also \citet{von11b}. Planets b, c, and e (the transiting super-Earth) are well inside the system's HZ (right panel). For details, see \S \ref{sec:hz} and Table \ref{tab:equiltemp}.}
  \label{fig:fig4}
\end{figure*}

Our calculations of the inner and outer boundaries of 55~Cnc's HZ are based on our directly determined host star properties (\S \ref{sec:properties}). Similar to \citet{von11} for GJ~581, we use the equations in \citet{und03} and \citet{jon10} to relate inner and outer edges of the HZ to the luminosity and effective temperature of the host star 55~Cnc. We find an inner and outer HZ boundary of 0.67~AU and 1.32~AU, respectively. The HZ is shown as the gray-shaded region in Figure \ref{fig:fig4}, which illustrates the architecture of the 55~Cnc system at different spatial scales. 

We calculate equilibrium temperatures $T_{eq}$ for the five known 55~Cnc planets using the equation
\begin{equation}\label{eq:equitemp}
  T_{eq}^4 = \frac{S (1 - A)}{f \sigma},
\end{equation}
where $S$ is the stellar energy flux received by the planet, $A$ is the Bond albedo, and $\sigma$ is the Stefan-Boltzmann constant \citep{skl07}. The energy redistribution factor $f$ indicates the atmospheric efficiency of redistributing the radiation received from the parent star across the planetary surface by means of circulation, winds, jet streams, etc. $f$ is set to 2 for a hot dayside (no heat redistribution between day and night side) and to 4 for even heat distribution\footnote{See Appendix in \citet{spi10} for a detailed explanation of the $f$ parameter.}. Table \ref{tab:equiltemp} shows the calculated equilibrium temperatures for the planets in the 55~Cnc system assuming different values of Bond albedos, inluding the value for Earth ($A=0.29$). Note that the temperature given for the $f=2$ scenario is the planet dayside temperature. All of 55~Cnc's known planets, except planet f, are either located well inside or beyond the system's HZ (see Figure \ref{fig:fig4}).

Figure \ref{fig:fig4} shows that planet 55~Cnc f \citep[$M \sin i = 0.155  M_{Jupiter} = 49.3 M_{\earth}$; table 10 in][]{daw10} is on an elliptical orbit ($e \simeq 0.3$) during which it spends about 74\% of its orbital period of approximately 260 days inside the HZ. Thus, $T_{eq}$ is a function of time (or phase angle). The time-averaged distance between planet f and 55~Cnc is 0.82 AU. For the $f=2$ scenario (no heat redistribution between day and night sides) and $A=0.29$, planet f's time-averaged dayside temperature is 294 K, and varies between 263 K (apastron) and 359 K (periastron) during its elliptical orbit. Assuming an even heat redistribution ($f=4$) and $A=0.29$, however, we calculate planet f's time-averaged surface temperature $T_{eq}^{f=4} = 247$ K, with a variation of 221~K at apastron to 302~K at periastron. Planet f's long-period, elliptical orbit  makes any kind of tidal synchronization unlikely. Further taking into account its mass, typical of gas giant planets, $f=4$ appears to be a much more likely scenario than $f=2$ or similar.
Note, that even though planet f's time-averaged $T_{eq}^{f=4}$ is below the freezing point of water, heating due to greenhouse gases could moderate temperatures in its atmosphere to  above the freezing point of water \citep[e.g.,][]{skl07, wor10}.



We point out that eccentricity estimates for planet f range from values between 0.13 and 0.3 in \citet{daw10} for the correct 0.74-day period of planet e with 1-$\sigma$ error bars of $\sim$0.05.  Clearly, the differences between the temperatures at apastron and periastron calculated above decrease if $e_f$ is smaller than our value of 0.3.  In addition, the orbit-averaged flux $S$ (Equation \ref{eq:equitemp}) increases with the planetary eccentricity $e$ as $\left(1-e^2\right)^{-1/2}$.  Thus, the equilibrium temperature of planet f increases for larger $e_f$, but this is at most a 10\% effect for $e_f < 0.4$.  Finally, irrespective of the instantaneous value of $e_f$, all the planets in the 55~Cnc system undergo long-term secular oscillations in eccentricity such that their effective temperatures may change in time.  

We further note that the planet 55~Cnc f is likely too massive to harbor liquid water on any
planetary surface \citep{skl07}.  In terms of an actual {\it habitable object} in the 55~Cnc system, there are thus two potential candidates: a massive moon in orbit around planet f or an additional low-mass planet in or near the HZ.  

Could 55~Cnc f host a potentially habitable moon?  One formation model suggests that the mass of a giant planet satellite should generally be about $10^{-4}$ times the planet mass \citep{can06}.  For 55~Cnc f the expected satellite mass is therefore $\sim 5 \times 10^{-3} M_\oplus$, which is comparable to the mass of Jupiter's moon Europa but probably too low to retain a thick atmosphere for long time scales at HZ temperatures \citep{wil97}.  Of course, there exist alternate origin scenarios that could produce more massive giant planet moons \citep[e.g.,][]{agn06} with correspondingly higher probability of atmospheric retention. In addition, values of $e_f < 0.3$ would slightly decrease the equilibrium temperature of any hypothetical moon around planet f, and consequently increase its potential of retaining an atmosphere. 

Any potential, dynamically stable existence of an additional low-mass planet in the outer part of 55~Cnc's HZ depends in large part on the eccentricity of planet f, since its gravitational reach will cover a larger fraction of the HZ for higher values of $e_f$.  Using the simple scaling arguments of \citet{jon05}, which are based on the analytical two-planet Hill stability limit \citep{mar82,gla93}, the  semimajor axis of the orbit closest to planet f that is likely to be stable for a low-mass planet is roughly 0.87 / 1.0 / 1.14 / 1.24 / 1.32 AU for $e_f = $ 0 / 0.1 / 0.2 / 0.3 / 0.4.  Given our calculated value of 1.32 AU for the outer edge of the HZ and our default value of $e_f = 0.3$, only the outermost portion of the HZ is able to host an additional low-mass planet. 

This simple scaling argument matches up reasonably well with the N-body simulations of \citet{ray08}, who mapped the stable zone of 55~Cnc between planets f and d but assumed a low fixed (initial) value for $e_f$. The exterior 3:2 mean motion with planet f, located at 1.02-1.04 AU, represents an additional stable niche. \citet{ray08} showed that, although narrow, this resonance is remarkably stable for long time scales, for planet masses up to the mass of planet f or even larger, and even for some orbits that are so eccentric that they cross the orbit of planet f.  This resonance is stable whether planet f's orbit is eccentric or circular.  

Thus, we conclude that if $e_f$ is large ($\gtrsim 0.3$) an additional Earth-like planet could exist in either the 3:2 resonance with planet f at 1.03 AU or in the outer reaches of the HZ.  In that case the additional planet's orbit would likely also be eccentric, although climate models have shown that large eccentricities do not preclude habitable conditions \citep{wil02,dre10}.  If, however, $e_f$ is small then an additional planet in HZ could be as close-in as about 0.9 AU and would likely be on a low-eccentricity orbit.


\begin{deluxetable*}{cllcccccc}
    \tablecaption{Equilibrium Temperatures for the 55~Cnc System Planets\label{tab:equiltemp}}
	\tablewidth{0pc}
    \tablehead{
    \colhead{} & & & \multicolumn{2}{c}{$A=0.1$} & \multicolumn{2}{c}{$A=0.29$} & \multicolumn{2}{c}{$A=0.5$} \\
    \cline{4-5} \cline{6-7} \cline{8-9} \\
      \colhead{Planet} & \colhead{$M_{\sin i}$ ($M_{Jup}$)} & \colhead{$a$ (AU)} & \colhead{$T_{eq}^{f=4}$ (K)} & \colhead{$T_{eq}^{f=2}$ (K)} & \colhead{$T_{eq}^{f=4}$ (K)} & \colhead{$T_{eq}^{f=2}$ (K)} & \colhead{$T_{eq}^{f=4}$ (K)} & \colhead{$T_{eq}^{f=2}$ (K)}}
      \startdata
     b & 0.825(3) & 0.1148(8) & 		$699 \pm 4$\phn\phn & 	$831 \pm 5$\phn\phn & 	$659 \pm 4$\phn\phn & 				$784 \pm 5$\phn\phn & 		$604 \pm 4$\phn\phn & 		$718 \pm 5$\phn\phn \\
     c & 0.171(4) & 0.2403(17) & 		$483 \pm 3$\phn\phn & 	$575 \pm 4$\phn\phn & 	$456 \pm 3$\phn\phn & 				$542 \pm 4$\phn\phn & 		$417 \pm 3$\phn\phn & 		$496 \pm 4$\phn\phn \\
     d & 3.82(4)  & 5.76(6) & 			$99 \pm 1$\phn & 		$117 \pm 1$\phn\phn & 	$93 \pm 1$\phn & 					$111 \pm 1$\phn\phn & 		$85 \pm 1$\phn & 			$101 \pm 1$\phn\phn \\
     e & 0.0260(10) & 0.01560(11) & 	$1895 \pm 12$\phn\phn & $2253 \pm 14$\phn\phn & $1786 \pm 12$\phn\phn & 	$2124 \pm 14$\phn\phn & 	$1636 \pm 12$\phn\phn & 	$1945 \pm 14$\phn\phn \\
     f & 0.155(8) & 0.781(6) & 			$268 \pm 2$\phn\phn & 	$319 \pm 2$\phn\phn & 	$253 \pm 2$\phn\phn & 				$301 \pm 2$\phn\phn & 		$231 \pm 2$\phn\phn & 		$275 \pm 2$\phn\phn \\
      \enddata

\tablecomments{Equilibrium temperatures for different values of the Bond albedo $A$, and based on the equations in \citet{skl07}. $A = 0.29$ corresponds to Earth's albedo. Planet masses and orbital element values are from table 10 in \citet{daw10}; note that 55~Cnc e's mass is not subject to a $\sin i$ uncertainty. $f = 4$ implies perfect energy redistribution efficiency on the planetary surface, $f = 2$ means no energy redistribution between day and night sides. For details, see \S \ref{sec:hz} and Figure \ref{fig:fig4}. The  uncertainties in $T_{eq}$ are based on the uncertainties in 55~Cnc's luminosity and the planets' orbital elements. Planet f's elliptical orbit ($e \simeq 0.3$) causes it to spend around 74\% of its year in 55~Cnc's circumstellar HZ (Figure \ref{fig:fig4}). For details, see \S \ref{sec:hz}.
}
\end{deluxetable*}


\section{The Transiting Super-Earth 55~Cnc e}\label{sec:planet}

This Section contains the implications of our calculated stellar parameters for the radius and bulk density of the transiting super-Earth 55~Cnc e. 

Three recent papers rely on the value of 55~Cnc's physical stellar radius that we measure to be $0.943 \pm 0.010 R_{\odot}$ (see \S \ref{sec:radius} and Table \ref{tab:properties}). \citet{win11} and \citet{dem11} both independently report the photometric detection of a transit of 55~Cnc e. In addition, \citet{kan11} calculate combinations of expected planetary brightness variations with phase, for which knowledge of planetary radius is useful. 

In our calculations of the planetary radius, we assume that 55~Cnc e has a cold night side, i.e., the planet is an opaque spot superimposed onto the stellar disk during transit. We note that, strictly speaking, any calculated planetary radius therefore actually represents a lower limit, but due to the intense radiation received by the parent star at this close proximity, it is unlikely for 55~Cnc e to retain any kind of atmosphere \citep{win11}. Using the formalism in \citet{win10}, the measured flux decrement during transit therefore corresponds to $(\frac{R_p}{R_\star})^2$, where $R_p$ and $R_{\star}$ are planetary and stellar radius, respectively. 

\citet{win11} obtain $\frac{R_p}{R_\star}$ = $0.0195 \pm 0.0013$ and $M_{p}$ = $8.63 \pm 0.35 M_{\Earth}$. In combination with our stellar radius, these measurements result in $R_{p} = 2.007 \pm 0.136 R_{\Earth}$ and a planetary bulk density of $5.882 \pm 0.728$ g cm$^{-3}$ or $1.067 \pm 0.132 \rho_{\Earth}$.

\citet{dem11} measure $\frac{R_p}{R_\star} = 0.0213 \pm 0.0014$ and assume a planetary mass of $7.98 \pm 0.69 M_{\Earth}$. Together with our radius for 55~Cnc, they imply a planetary radius of $2.193 \pm 0.146 R_{\Earth}$ and a bulk density of $4.173 \pm 0.602$ g cm$^{-3}$, corresponding to $0.757 \pm 0.109 \rho_{\Earth}$\footnote{We note that we are calculating uncertainties based on simple Gaussian error propagation. That is, we make the assumption that the errors are not correlated, which is not quite correct since, e.g., stellar mass can be related to stellar radius.}. 

We refer the reader to the planet transit discovery papers, in particular figure 3 in \citet{win11} and figures 5 \& 6 in \citet{dem11}, for comparison of these density values to other transiting super-Earths in the literature. It should be noted that, although 55~Cnc is the brightest known star with a transiting planet, the amplitude of the transit signal is among the smallest known to date, making the determination of planet radius very difficult.





\section{Summary and Conclusion}    \label{sec:conclusion}

Characterization of exoplanets is taking an increasingly central role in the overall realm of planetary research. An often overlooked aspect of determining the characteristics of extrasolar planets is that reported physical quantities are actually dependent on the astrophysical parameters of the respective host star, and that assumptions of varying degree of certainty may implicitly be contained in quoted absolute values of exoplanet parameters. Consequently, the necessity of ``understanding the parent stars'' can hardly be overstated, and it served as the principal motivation for the research presented here. 

Our new interferometric measurements provide a directly determined, high-precision angular radius for the host star 55~Cnc. We couple these measurements with trigonometric parallax values and literature photometry to obtain the stellar physical diameter ($0.943 \pm 0.010R_{\odot}$), effective temperature (5196 $\pm$ 24 K), luminosity ($0.582 \pm 0.014L_{\odot}$), and characteristics of the HZ (see Table \ref{tab:properties}). This shows that planet f spends 74\% of its year in the HZ. We use isochrone fitting to calculate 55~Cnc's age (10.2 $\pm$ 2.5 Gyr) and mass ($0.905 \pm 0.015M_{\odot}$). Finally, the directly determined stellar radius allows for a model-independent estimate of the radius of any transiting extrasolar planets. We use our stellar diameter value and recently published numbers for $\frac{R_p}{R_\star}$ to estimate the radius ($\sim 2.05 \pm 0.15R_{\earth}$) and bulk density (0.76 or 1.07 $\rho_{\earth}$, depending on the assumed planetary mass) of the transiting planet 55~Cnc e. 

Due to its (naked-eye) brightness and consequent potential for detailed spectroscopic studies, the small size of the transiting super-Earth 55~Cnc e, planet f's location in the circumstellar HZ, and generally the fact that 55~Cnc hosts at least five planets at a wide range of orbital distances, the system will undoubtedly be the source of exciting exoplanet results in the very near future.



\acknowledgments

The authors would like to thank B.-O. Demory and J. N. Winn for many open conversations and exchange of extremely useful information about the planetary radius of 55~Cnc e during the preparation of this manuscript, E. K. Baines for discussions about interferometric data quality, and D. Spiegel for very helpful suggestions on the issues interferometric visibilities and planet habitability. We would further like to extend our gratitude to the anonymous referee for her/his careful reading of the manuscript and the insightful comments that improved the quality of this publication. TSB acknowledges support provided by NASA through Hubble Fellowship grant \#HST-HF-51252.01 awarded by the Space Telescope Science Institute, which is operated by the Association of Universities for Research in Astronomy, Inc., for NASA, under contract NAS 5-26555. STR acknowledges partial support from NASA grant NNH09AK731. The CHARA Array is funded by the National Science Foundation through NSF grants AST-0606958 and AST-0908253 and by Georgia State University through the College of Arts and Sciences, the W. M. Keck Foundation, the Packard Foundation, and the NASA Exoplanet Science Institute. This research made use of the SIMBAD literature database, operated at CDS, Strasbourg, France, and of NASA's Astrophysics Data System. This publication makes use of data products from the Two Micron All Sky Survey, which is a joint project of the University of Massachusetts and the Infrared Processing and Analysis Center/California Institute of Technology, funded by the National Aeronautics and Space Administration and the National Science Foundation. This research made use of the NASA/IPAC/NExScI Star and Exoplanet Database, which is operated by the Jet Propulsion Laboratory, California Institute of Technology, under contract with the National Aeronautics and Space Administration.




\bibliographystyle{apj}            

\bibliography{apj-jour,paper}      

\end{document}